\documentclass{llncs}
\usepackage[dvips]{graphicx}
\usepackage{amsmath}
\usepackage{amssymb}
\usepackage{verbatim}
\usepackage{textcomp}
\usepackage{amsfonts}

\newtheorem{observation}{Observation}
\newtheorem{fact}{Fact}

\author{M. V. Panduranga Rao}
\institute{Department of Computer Science and Automation \\Indian Institute of Science\\ Bangalore  560 012\\ India.
\\pandurang@csa.iisc.ernet.in}
\date{}
\begin{document}
\newlength{\twidth}
\title{Generalized Counters and Reversal Complexity}
\maketitle
\begin{abstract}
We generalize the definition of a counter and counter reversal
complexity and investigate the power of generalized deterministic 
counter automata in terms of language recognition. 
\end{abstract}

\section{Introduction}
Deterministic  counter (DC) automata are essentially deterministic finite automata (DFA) enhanced with counters. A (conventional) counter 
is a device capable of storing an integer on which three operations can be performed by the finite control: 
increment, decrement (or do nothing) and test-for-zero.  The input lies on a tape demarcated by end-markers ``\textcent" and ``$\$$", and is read by
a read-only head. 

Two counters can simulate a tape, and therefore a two-counter machine is as 
powerful as a Turing machine. However, imposing restrictions on resources
yields proper subclasses of recursive languages.
A deterministic counter machine might be restricted in terms of different 
resources like (a) the number of counters it is 
equipped with, (b) if the head can move both ways on the input tape 
(if so, how many such \emph{head reversals} are allowed) and 
(c) the number of times the counter(s) is allowed to switch between increment and decrement modes (called \emph{counter reversal complexity}).
Moreover, restrictions may also be imposed on the types of actions 
permitted.
\emph{Blind} counter machines have no information available on 
the contents and sign of the counters. \emph{Partially blind} counter machines
are also blind; in addition, the contents of the counter must always be non-negative.
The machine crashes on being driven below zero~\cite{Greibach,Say}.

In this paper we generalize the notion of a counter and investigate the resulting increase in power.
Generalizations based on group theory have been  proposed and investigated by various authors \cite{DM,MS1,MS2}. 
Instead of a simple counter as described above, the finite automaton is equipped with a group on which it can perform the group operation.
The counter can store any element of the group. 
However, the exact element of the group currently contained in the counter is not available to the finite control: it can only check whether
it is the identity element or not. The power of various groups (abelian, non-abelian, free etc.) has been studied extensively in the above mentioned
papers.

The main contribution of this paper is threefold.
First, we propose a generalized algebraic structure for counters that includes a notion of ``negativeness"  
(in section 2, along with some preliminaries). In the process, we introduce
a generalized notion of counter reversal complexity.
Secondly, we examine specific instances\footnote{In this paper, we consider machines that have only one counter.
Further, unless otherwise stated, the machines will be partially blind and accept by final state and 
empty store.}
of the generalized counter and show that they recognize non-trivial languages with low counter 
and head reversal complexity and overall time complexity. 
Duris and Galil \cite{DG} showed a witness language that 
cannot be recognized by any 2-way deterministic one-counter machine,
while it can be recognized by a 2-way deterministic pushdown automata (2DPDA) with one
stack.
We show that a powerful instance of the generalized counter given in this paper can recognize this language with 
small counter as well as head reversal complexity (in section 3).
And finally, we establish a hierarchy among the corresponding 1-way versions in terms of language recognition in section 4. Section 5 concludes 
the paper.

\section{Related Previous Work and Definition of a Generalized Counter}
We first give a formal definition of 2-way one-counter deterministic (2DC) 
automata.

\begin{definition}
A 2DC machine $M$ is a 5-tuple $ ( Q,\Sigma,q_0,\delta,F)$ where $Q$ is a finite set of states, $q_0$ a special start state,
$F\subseteq Q$ the set of accepting states and 
$\Sigma$ is a finite input alphabet. $\delta$ is a mapping from $Q\times (\Sigma \bigcup \{$\textcent$,\$\})\times\{0,1\}$ to 
 $Q\times \{-1,0,+1\}\times\{-1,0,+1\}$.
 $\square$
\end{definition}

The transition function takes three input parameters: the current state, the current symbol being read, and the status of the counter (say, $0$ if the counter reads zero and 
$1$ if non-zero), and does the following: changes the state, moves the head by $-1$, $0$ or $+1$ position on the tape, and changes the 
counter value by $-1$, $0$ or $+1$. 

If the machine is blind (2BDC), the transition function does not get any 
information from the counter. 
Transitions depend only on the current state
and the symbol being scanned. The machine accepts by final state and empty counter.
A partially blind machine (2PBDC) crashes if the counter 
goes negative at any stage in the computation.

Many results regarding the power of various models of counter machines
exist~\cite{Greibach,GI,IT,ISK,Say,Say1,Mon,Pet}.

We now give a formal definition of our abstract generalized counter.

\begin{definition}
Consider a group $(U,\circ)$. Let $G=\{A_1, \ldots, A_{k}\}\subset U$ be a finite \emph{counter generating set}
and $G_{inv} = \{X \in U \mid X^{-1} \in G\}$ such that $G^*\bigcap G_{inv}^* =\phi$ 
where ``*" denotes the closure operation and $\phi$ is the null set. 
Let $F_- \subset U$ be such that $F_- \bigcap G^* = \phi$, $G_{inv}^* \subseteq F_-$
and membership in $F_-$ is decidable in constant time\footnote{Checking if an element in the additive (multiplicative) group of 
real numbers is $\geq 0$ ($\geq 1$) is an example of such a membership test. The counter is endowed with the capability of performing such tests.}. 
Let $F_+ = U\backslash F_-$.

We call the tuple $(U,G,F_-)$ a \emph{generalized counter}.
Then, at any step $t$,  $\Omega_t \in F_+$ serves as the \emph{non-negativity} 
condition.

Define the operation
$increment(i)$ on $\Omega_{t-1}$ to be 
$X_i \Omega_{t-1} = \Omega_t$ and 
$decrement(i)$ on $\Omega_{t-1}$ to be $ X_i^{-1} \Omega_{t-1} = \Omega_t$ for $X_i \in G$.
In general, $X_iX_j\neq X_jX_i$, for $X_i,X_j\in G\bigcup G_{inv}$, $i\neq j$.
We uniformly identify an incoming $X$ at step $t$ as a left operand.
$\square$
\end{definition}

Observe that in this setting, conventional counter is  $C_\mathbb{Z}=(\mathbb{Z},\{1\},\mathbb{Z}^-)$.
We now give the formal definition of a deterministic automaton with a generalized counter.

\begin{definition}
A 2DC($(U,G,F_-)$) machine $M$ is a 6-tuple $ ( Q,\Sigma,q_0,\delta,F,(U,G,F_-))$ where $Q$ is a finite set of states, $q_0$ a special start 
state, $F\subseteq Q$ the set of accepting states and 
$\Sigma$ is a finite input alphabet. $\delta$ is a mapping from $Q\times (\Sigma \bigcup \{$\textcent$,\$\})\times\{0,1\}$ to 
 $Q\times \{-1,0,+1\}\times (G\bigcup G_{inv})$.
 $\square$
\end{definition}

In case of partially blind counter machines, the transition function behaves as 
follows: $\forall q\in Q$ and $\sigma\in \Sigma \bigcup \{$\textcent, $\$\}$, 
(a) $\delta(q,\sigma,\Omega_1)= \delta(q,\sigma,\Omega_2)$ for all $\Omega_1,\Omega_2\in F_+$ (blindness)
and (b) $\delta(q,\sigma,\Omega)=\phi$ if $\Omega\in F_-$ (non-negativity). 

Note that $F_-$ is not relevant in 2DC machines that are allowed to store
negative elements. It is important, however, for specifying partially blind 
machines. We include it in all models, for uniformity of presentation.
We call a 2-way machine with a generalized counter $(U,G,F_-)$ a 2DC($(U,G,F_-)$) machine and the class of languages
recognized by such machines, $\mathcal{L}$(2DC($(U,G,F_-)$)). 
Similar conventions will be followed for PBDC machines also.
\section{Applications}
We now discuss two powerful instances of the abstract counter defined in the previous section. 

\subsection{A Counter Over Reals}

Consider the counter $C_{\mathbb{R}}(k) = (\mathbb{R},\{\rho_1, \ldots, \rho_{k}\},\mathbb{R}^- )$,
where $\mathbb{R}$ is the additive group of real numbers, $\rho_1,\ldots,\rho_{k}$ are square roots of distinct prime numbers for some constant 
$k$ and $\mathbb{R}^-$ is the set of negative reals.
The non-negativity condition is therefore, $\Omega_t \in \mathbb{R}^+ \bigcup 0$.

It is easy to see that this counter is at least as powerful as the conventional counter.
We now prove that machines with a $C_{\mathbb{R}}(k)$ counter can do more.
The language $L_{abc}= a^nb^nc^n$ is context sensitive and is therefore not recognizable by any 1DPDA. We can show that there exists an  
algorithm to recognize the general family of such languages using a $C_{\mathbb{R}}(k)$ counter.

\begin{theorem} 
There exists a 1PBDC($C_{\mathbb{R}}(k-1)$) machine that recognizes
$L_{gen}= \{a_0^{n}a_1^{l_1n}\ldots a_{k-1}^{l_{k-1}n}\mid n\in \mathbb{N}\}$,
where $a_0,a_1,\ldots, a_{k-1}$ are symbols of a finite alphabet and 
$l_i\in \mathbb{N}$, with one counter reversal and no head reversal.
\end{theorem}
\begin{proof}
We begin by noting the following.

\begin{definition}
A set of $n$ real numbers $\alpha_1,\ldots,\alpha_n$ is said to be rationally dependent if the relation
$c_1\alpha_1+\ldots+c_n\alpha_n=0$ holds for some rational numbers $c_1,\ldots,c_n$, not all zero.
A set that is not rationally dependent is said to be rationally independent.
$\square$
\end{definition}

\begin{fact}
Any set of square roots of distinct prime numbers is rationally independent.
\end{fact}

We use square roots $\rho_1,\ldots,\rho_{k-1}$ of $k-1$ distinct primes.
The  1PBDC($C_{\mathbb{R}}(k)$) machine works as follows. That the input $x$ is indeed of the form
$a_0^*a_1^*\ldots a_{k-1}^*$ is verified by the DFA as the input is scanned. On scanning an $a_0$,
the counter is incremented by
$(l_1\rho_1 +\ldots + l_{k-1}\rho_{k-1})$.
Hence, after having scanned all the $a_0$'s, the counter holds $(l_1\rho_1+\ldots+ l_{k-1}\rho_{k-1})n_0$, for some $n_0\in \mathbb{Z}^+$.

As the head moves further, on scanning an $a_i$, $1\leq i\leq k-1$, the counter is decremented by $\rho_i$.

The counter holds $0$ if and only if
\[
(l_1\rho_1 +\ldots + l_{k-1}\rho_{k-1} )n_0 =  n_1\rho_1 +\ldots + n_{k-1}\rho_{k-1}
\]
where $n_1, \ldots, n_{k-1}$ are the number of $a_1,\ldots, a_{k-1}$ symbols respectively in the input string.
Since $\rho_1,\ldots,\rho_{k-1}$ are rationally independent by fact 1, the above equation is true if and only if
$l_1n_0=n_1,\ldots, l_{k-1}n_0=n_{k-1}$. In other words, the counter
reads $0$ if and only if the input is in the language. $\square$
\end{proof}

In the next section, we will show limitations of the real counter in spite of having the facility of arbitrary precision.
\subsection{A Matrix Counter}

The operands of a general matrix counter are finite dimensional invertible matrices, the 
operator being (left) matrix multiplication. 
The matrix counter is defined by $(GL(m,\mathbb{R}),\{A_1,\ldots, A_{k}\}, F_-)$
where $GL(m,\mathbb{R})$ is the group of $m$-dimensional invertible matrices over $\mathbb{R}$, and
$F_- = \{X\in GL(m,\mathbb{R})\mid |X| < 1\}$ where
$|.|$ is defined as:
\[
|X| = \sqrt{\sum_i\sum_j|X_{ij}|^2}.
\]
Therefore, the non-negativity condition is 
$\sqrt{\sum_i\sum_j |(\Omega_t)_{i,j}|^2} \geq 1$
at any time $t$ during the computation.
\begin{theorem}
$\mathcal{L}$(2PBDC($C_\mathbb{R}(k)$)) $\subseteq$ $\mathcal{L}$(2PBDC($C_\mathbb{M}(k)$)).
\end{theorem}
\begin{proof}
Given any 2PBDC($C_\mathbb{R}(k)$) machine $M$ we construct a 2PBDC($C_\mathbb{M}(k)$) machine $M'$ that
recognizes the same language as $M$ as follows.

Suppose $M$ is described by the tuple $(Q,\Sigma, q_0, F,\delta,C_\mathbb{R}(k))$. Then, $M'$ is
$(Q,\Sigma, q_0, F,\delta',C_\mathbb{M}(k))$. The matrix counter and  $\delta'$ are defined as follows.
Suppose the generating set of $C_\mathbb{R}(k)$ is $\{\rho_1,\ldots,\rho_{k} \}$,
consisting of square roots of distinct primes as mentioned earlier.

Then the $C_\mathbb{M}(k)$ counter is 
$\big(GL(1,\mathbb{Z}),\{[p_1],\ldots,[p_{k}]\}, F_- \big)$
where $p_i$, $1\leq i\leq k$, are the primes $\rho^2$ and $F_-=\{ [x] \mid |[x]|<1\}$. 
If $\delta(q,\sigma, \beta) = (q',D, \rho_i \textrm{ or } -\rho_i)$, then $\delta'(q,\sigma, \beta) = (q',D, [p_i]\textrm{ or }[p_i]^{-1})$,
where $q,q'\in Q$, $\sigma\in \Sigma\bigcup \{\textrm{\textcent},\$\}$, $\beta \in \{0,1\}$, $D\in \{-1,0,+1\}$,
$0\leq i\leq k$. 
Thus, if the real counter holds 
$(n_{a,1} - n_{b,1})\rho_1 +\ldots + (n_{a,k} - n_{b,k})\rho_{k}$, the matrix counter
holds $[p_1^{n_{a,1} - n_{b,1}}p_2^{n_{a,2} - n_{b,2}}\ldots p_{k}^{n_{a,k} - n_{b,k}}]$.
Therefore, the matrix counter contains $[1]$ if the real counter contains $0$.
Further, as long as the content of the real counter is greater than $0$, 
the non-negativity condition is maintained for the matrix counter also, because of
the way in which the primes have been chosen.
$\square$
\end{proof}

Duris and Galil \cite{DG} showed that no 2DC can 
recognize 

$L_{pat}=\{x_0\#\ldots\#x_k\# \mid k\geq 1, x_j\in\{0,1\}^*
\textrm{ for } 0\leq j\leq k, \textrm{ for some } 1\leq i\leq k,\textrm{ } x_i=x_0\}$,
where a substring between two successive $\#$'s is called a block.
In this section we show a matrix counter machine that recognizes $L_{pat}$.  

\begin{theorem}
There exists a 2DC($C_{\mathbb{M}}(k)$) machine that recognizes $L_{pat}$. The number of reversals $O(m)$ where $m$ is the number of
blocks in the input.
\end{theorem}
\begin{proof}
We use a theorem of Ambainis and Watrous:
\begin{theorem}[Ambainis and Watrous \cite{AW} ] Let 
\begin{displaymath}
\mathbf{A} = 
\left( \begin{array}{ccc}
4 & 3 & 0 \\
-3 & 4 & 0 \\
0 & 0 &  5 \\
\end{array} \right)
\textrm{ and } \mathbf{B} = 
\left( \begin{array}{ccc}
4 &  0 & 3 \\
0  & 5 & 0 \\
-3 & 0 &  4 \\
\end{array} \right)
\end{displaymath}
and $u$ be a $3\times 1$ vector with components $u[1]$, $u[2]$ and $u[3]$.

Let $u = Y_1^{-1}\ldots Y_n^{-1} X_n\ldots X_1 (\textrm{1 0 0 })^T$ where
$X_j,Y_j\in \{A,B\}$. 
Then, $u[2]^2 + u[3]^2=0$ if and only if $X_j=Y_j$ for $1\leq j\leq n$.
$\square$

\end{theorem}
Observe that 
\begin{displaymath}
\mathbf{A^{-1}} = 
\frac{1}{25}\left( \begin{array}{ccc}
4 & -3 & 0 \\
3 & 4 & 0 \\
0 & 0 &  5 \\
\end{array} \right)
\textrm{ and } \mathbf{B^{-1}} = 
\frac{1}{25}\left( \begin{array}{ccc}
4 &  0 & -3 \\
0  & 5 & 0 \\
3 & 0 &  4 \\
\end{array} \right)
\end{displaymath}
Let the counter be $(GL(3,\mathbb{R}),\{A,B\},F_-) $ where $F_-$ is defined as before.

In this proof, ``scanning" a block (in whichever direction) is also meant to involve a non-trivial operation (i.e. other than $I$, the identity matrix) on the 
counter for every symbol in the block.
We give a 2DC($C_{\mathbb{M}}(k)$)  algorithm that recognizes $L_{pat}$. For the sake of clarity in presenting the algorithm, we first define
two ``subroutines":

subroutine \begin{bf}increment\end{bf} $\quad\quad\quad \quad\quad\quad\quad $ subroutine \begin{bf}decrement\end{bf}

$\quad$  If a ``0" is being scanned $\quad \quad\quad\quad\quad\quad $ If a ``0" is being scanned

$\quad \quad \quad$ $\Omega_{t+1}:= A\Omega_t$. $\quad\quad\quad \quad\quad\quad \quad \quad\quad \quad$ $\Omega_{t+1}:= A^{-1}\Omega_t$.

$\quad$ { If a ``1" is being scanned }  $\quad\quad\quad\quad \quad\quad$ If a ``1" is being scanned

$\quad \quad  \quad$ $\Omega_{t+1}:= B\Omega_t$. $\quad\quad\quad\quad\quad\quad\quad\quad \quad \quad$  $\Omega_{t+1}:= B^{-1}\Omega_t$.

 where $\Omega_t$ is the content of the counter at step $t$.
The algorithm is as follows:

Initially $\quad$$\Omega_0$=I.

$\quad \quad$ Until the first ``$\#$" is encountered,

$\quad \quad \quad$ scan right from \textcent~performing \begin{bf}increment\end{bf}.

$\quad \quad$  For all subsequent blocks do: 

$\quad \quad \quad$ scan from the \emph{right} ``$\#$" to that on the left, performing \begin{bf}decrement\end{bf}.

$\quad \quad \quad \quad $  if $\Omega_t=I$  \begin{bf}accept\end{bf}.

$\quad \quad \quad $   scan from the left ``$\#$" to that on the right, performing \begin{bf}increment\end{bf}.

$\quad \quad \quad $   move to the next block.

$\quad$   \begin{bf}reject\end{bf}.

Let $C_{x}$ stand for the product of the matrices taken from $G$ 
applied while scanning a block $x$ in the forward direction.
Similarly, let $C_{x}^{-1}$ be the product of the matrices taken from $G_{inv}$,
applied while scanning a block $x$ in the reverse order.

The 2DC($C_{\mathbb{M}}(k)$) machine $M$ recognizes $L_{pat}$ as follows.
Initially the counter contains the identity matrix $I$.
After scanning $x_0$, let the counter contain $C_{x_0}$. 
For every subsequent block $i$, it checks if $C_{x_i}^{-1}C_{x_0}=I$.
This will be the case if and only if $x_0=x_i$, by theorem 4.
If $C_{x_i}^{-1}C_{x_0}\neq I$, the matrices applied in
the current block are undone while scanning to the $\#$ on the right
end of the block so that the counter contains
$C_{x_i}C_{x_i}^{-1} C_{x_0} = C_{x_0}$ just before entering 
the next block. 

Since there are only two reversals of the counter per block, 
the reversal complexity of the algorithm is $O(m)$ where $m$ is
the number of blocks in the input string.
$\square$
\end{proof}

\section{One-Way Versions}
In this section we discuss some results regarding 1-way PBDC automata.

Let us first note some useful facts.
One can view the counter as a container into which marked coins are added or taken out.
Incrementing or decrementing the counter by $X_i$ corresponds to putting a coin marked $X_i$ into the counter or taking it out respectively,
satisfying the non-negativity condition at any given time. Therefore,

\begin{observation}
A counter can hold only countably many values. $\square$
\end{observation}

The following is an immediate consequence.

\begin{lemma}
Let $M=(Q,\Sigma,q_0,\delta,F,C_\mathbb{R}(k))$ and $\Omega_x$ denote the state of the counter after having read a string $x\in \Sigma^*$. 
Then, if there exists a positive real $\alpha$ such that $\Omega_x \leq \alpha$ for all $x\in\Sigma^*$, then $L(M)\in \mathcal{L}(REG)$, the class of
regular languages.
\end{lemma}
\begin{proof}
The above observation implies that in such a machine, the counter can contain only finitely many values. 
Therefore the ``state space" of the counter can be absorbed into the finite control itself, resulting in a DFA.
$\square$
\end{proof}

\begin{definition} If at any step, the 1PBDC(C) machine is in state $q\in Q$, the head is reading the first symbol of the $x$ and the counter
contains $\Omega$, then the triple ($q,x,\Omega$) describes its \emph{instantaneous configuration}.
$\square$
\end{definition}

If a 1PBDC(C) configuration ($q,x,c$) yields ($q',\epsilon,c'$), where $\epsilon$ denotes the empty string, after scanning $x$, then we write
$(q,x,\Omega) \models_x (q',\epsilon,\Omega')$.

We now state the main theorem of this section.

\begin{theorem}
$\mathcal{L}$(1PBDC($C_\mathbb{Z}$)) $\subsetneq$ $\mathcal{L}$(1PBDC($C_\mathbb{R}(k)$)) $\subsetneq$ $\mathcal{L}$ (1PBDC($C_\mathbb{M}(k)$)).
\end{theorem}
\begin{proof}
(a) $\mathcal{L}$(1PBDC($C_\mathbb{Z}$)) $\subsetneq$ $\mathcal{L}$(1PBDC($C_\mathbb{R}(k)$)):

The conventional counter over $\mathbb{Z}$ is a special case of the counter over reals. So, $\mathcal{L}$(1PBDC($C_\mathbb{Z}$)) $\subseteq$ 
$\mathcal{L}$(1PBDC($C_\mathbb{R}(k)$)).
Further,  by theorem 1, $L_{abc} \in \mathcal{L}$( 1PBDC($C_\mathbb{R}(k)$)). Since the conventional one-way counter machine is weaker
than pushdown automata which cannot recognize $L_{abc}$, it follows that  $L_{abc} \notin \mathcal{L}$(1PBDC($C_\mathbb{Z}$)).

(b)
$\mathcal{L}$(1PBDC($C_\mathbb{R}(k)$)) $\subsetneq$ $\mathcal{L}$(1PBDC($C_\mathbb{M}(k)$)):

That $\mathcal{L}$(1PBDC($C_\mathbb{R}(k)$)) $\subseteq$ $\mathcal{L}$(1PBDC($C_\mathbb{M}(k)$)) follows from theorem 2.
To prove proper containment, we need an ``interchange" lemma.
\begin{lemma}
Let $C_{\mathbb{R}}(k)$ be a real counter as defined in the previous section, 
with $\rho_k$ as the largest element in the generating set,
and let $L$ be a language in $\mathcal{L}$(1PBDC(($C_\mathbb{R}(k)$).
There is a constant $r$ and two integers $1\leq l < m \leq r$ such that for any decomposition of an input $x=v_1w_1v_2w_2\ldots v_rw_rv_{r+1} \in L$
with $\Omega_{v_1} \geq (\sum_{i=2}^r|v_i| + \sum_{i=1}^r|w_i|)\rho_k$, $|w_i|\geq 1$, we have that the string 
$x'=v_1w_1'v_2w_2'\ldots v_rw_r'v_{r+1}$ with $w_l'=w_m$, $w_m'=w_l$ and $w_i'=w_i$ for $i\notin\{l,m\}$, is also in $L$.

\end{lemma}
\begin{proof}
The proof proceeds on the lines of the interchange lemma in  \cite{MS1}.

Let $M=(Q,\Sigma,q_0,\delta,F,C_\mathbb{R}(k)$  be a 1PBDC($C_\mathbb{R}(k)$) machine. Let $r=|Q|^2+1$. Consider a string $x\in L(M)$,
and a decomposition $x= v_1w_1v_2w_2\ldots v_rw_rv_{r+1}$, $|w_i|\geq 1$, such that $\Omega_{v_1} \geq (\sum_{i=2}^r|v_i| + \sum_{i=1}^r|w_i|)\rho_k$. 
Then there exist $q_i,s_i\in Q$, $1\leq i\leq r$, $q_f\in F$ such that
\begin{eqnarray}
(q_{i-1},v_i,0) &\models_{v_i}& (s_i,\epsilon, \Omega_i),\textrm{  $1\leq i\leq r$}  \nonumber\\
(s_i,w_i,0) &\models_{w_i}& (q_i,\epsilon, \Omega'_i),\textrm{  $1\leq i\leq r$}  \nonumber \\
(s_r,v_{r+1},0) &\models_{v_{r+1}}& (q_f,\epsilon, \Omega_{r+1}). \nonumber
\end{eqnarray}
Since $x\in L$, $\Omega_1 + \Omega'_1 + \Omega_2 + \Omega'_2 + \ldots + \Omega_r+ \Omega'_r + \Omega_{r+1} =0$.
Since there are at most $|Q|^2$ pairs of tuples in $Q\times Q$, and the input has a length greater than $|Q|^2$, by the pigeon hole principle
we have $(s_l,q_l)=(s_m,q_m)$ for some $1\leq l <m \leq r$.

Now consider $x' =v_1w_1'v_2w_2'\ldots v_rw_r'v_{r+1}$ with $w_l'=w_m$, $w_m'=w_l$ and $w_i'=w_i$ for $i\notin\{l,m\}$.
Then,
\begin{eqnarray}
(q_{i-1},v_i,0) &\models_{v_i}& (s_i,\epsilon, \Omega_i),\textrm{  $1\leq i\leq r$}  \nonumber\\
(s_i,w'_i,0) &\models_{w'_i}& (q_i,\epsilon, \Omega''_i),\textrm{  $1\leq i\leq r$}  \nonumber \\
(s_r,v_{r+1},0) &\models_{v_{r+1}}& (q_f,\epsilon, \Omega_{r+1}) \nonumber
\end{eqnarray}
with $\Omega''_l=\Omega'_m, \Omega''_m=\Omega'_l$ and $\Omega''_i=\Omega'_i$ for $i\notin\{l,m\}$.
So,
$\Omega_1 + \Omega''_1 + \Omega_2 + \Omega''_2 + \ldots + \Omega_r+ \Omega''_r + 
\Omega_{r+1} = \Omega_1 + \Omega'_1 + \Omega_2 +\Omega'_2 + \ldots + \Omega_r+ \Omega'_r + \Omega_{r+1} =0$.
Note that $\Omega_1=\Omega_{v_1}$ has been chosen such that the interchange still satisfies the non-negativity condition. Therefore,
$x'$ also belongs to $L$.
$\square$
\end{proof}

Let $L_{pal}$ be $\{x\#x^R \mid x\in \Sigma^*\}$, where $x^R$ is the string $x$ reversed.

\begin{lemma}
Suppose $M$ is a $C_\mathbb{R}(k)$ counter machine recognizing $L_{pal}$. Then, for any positive $\alpha$, there exists a string $v_1\in \Sigma^*$
such that $\Omega_{v_1}> \alpha$.
\end{lemma}
\begin{proof}
Follows from lemma 1 and the fact that $L_{pal}$ is not regular.
$\square$
\end{proof}

\begin{lemma}
$L_{pal} \notin$ $\mathcal{L}$(1PBDC($C_\mathbb{R}(k)$)).
\end{lemma}
\begin{proof}
Let $L_{pal}$ be recognized by a 1PBDC($C_\mathbb{R}(k)$) machine $M$. 
Let $r$ be the constant from the interchange lemma.
Consider 
$x=v_1w_1v_2w_2\ldots\#\ldots w_{r-1}v_rw_rv_{r+1}$ in $L_{pal}$,
where $w_1\neq w_2$ and $\Omega_{v_1} \geq (\sum_{i=2}^r|v_i| + \sum_{i=1}^r|w_i|)\rho_k$.
By the previous lemma, such a $v_1$ exists.
Note that $v_{r+1}=v_1^R, w_r=w_1^R,v_r=v_2^R$ and $w_{r-1}=w_2^R$.
Then, by the interchange lemma, $x'= v_1w_2v_2w_1\ldots\#\ldots w_2^Rv_2^Rw_1^R        v_1^R$ also belongs to $L_{pal}$, a contradiction.
$\square$
\end{proof}

However, a simple modification of the algorithm to recognize $L_{pat}$ given in the previous section recognizes $L_{pal}$.
The tape head is now 1-way, and the counter is queried only on reading ``$\$$".
Therefore, $\mathcal{L}$(1PBDC($C_\mathbb{R}(k)$)) $\subsetneq$ $\mathcal{L}$(1PBDC($C_\mathbb{M}(k)$).
$\square$
\end{proof}

\section{Discussion}
In this paper we proposed a natural generalization of the counter. The generalization helps in analyzing the performance of
a counter machine in terms of reversal complexity of the counter. We established a hierarchy of counters when the head is 
restricted to move only forward. We believe that characterizing languages recognized by various types of counter machines
and their comparison with existing models are interesting problems to be addressed.

\bibliographystyle{plain}
                                       \bibliography{../../thesis/29_4_thesis/bib}

\end{document}